\documentclass[10pt]{article}
\usepackage [top=25truemm,bottom=25truemm,left=25truemm,right=25truemm]{geometry}
\usepackage{fancybox}
\usepackage{latexsym,amssymb,amsfonts}
\usepackage[dvips]{graphicx}
\usepackage{cite}
\usepackage{mysty}
\def\revise{}

\def\theequation{\arabic{equation}}

\newtheorem{proposition}{Proposition}
\newtheorem{defin.}[proposition]{Definition}
\newtheorem{theorem}[proposition]{Theorem}

\newtheorem{lemma}[proposition]{Lemma}

\title{\bf Directivity of quantum walk via\\ its random walk replica}

\author{Tomoki Yamagami $^{1,\,*}$
\and Etsuo Segawa $^2$
\and Nicolas Chauvet $^1$
\and Andr\'e R\"ohm $^1$
\and Ryoichi Horisaki $^1$
\and Makoto Naruse $^{1}$
}
\date{}

\begin{document}
\maketitle
\vspace{-2.5\baselineskip}
\begin{center}
{\small
$^1$ Department of Information Physics and Computing, Graduate School of Information Science and Technology, The University of Tokyo, 7-3-1 Hongo, Bunkyo, Tokyo 113-8656, Japan.\\
$^2$ Graduate School of Environment and Information Sciences, Yokohama National University, 79-1 Tokiwadai, Hodogaya, Yokohama, Kanagawa 240-8501, Japan.\\
$^*$ Email address: \texttt{yamagami-tomoki-qwb@g.ecc.u-tokyo.ac.jp}
}\vspace{1\baselineskip}\\
\end{center}
\noindent{\bf Abstract}\quad 
Quantum walks (QWs) exhibit different properties compared with classical random walks (RWs), most notably by linear spreading and localization. In the meantime, random walks that replicate quantum walks, which we refer to as quantum-walk-replicating random walks (QWRWs), have been studied in the literature where the eventual properties of QWRW coincide with those of QWs. However, we consider that the unique attributes of QWRWs have not been fully utilized in the former studies to obtain deeper or new insights into QWs. In this paper, we highlight the directivity of one-dimensional discrete quantum walks via QWRWs. By exploiting the fact that QWRW allows trajectories of individual walkers to be considered, we first discuss the determination of future directions of QWRWs, through which the effect of linear spreading and localization is manifested in another way. Furthermore, the transition probabilities of QWRWs can also be visualized and show a highly complex shape, representing QWs in a novel way. Moreover, we discuss the first return time to the origin between RWs and QWs, which is made possible via the notion of QWRWs. We observe that the first return time statistics of QWs are quite different from RWs, caused by both the linear spreading and localization properties of QWs.
\vspace{1\baselineskip}\\
\noindent{\small
$\blacktriangleright$  We use the following descriptions without notice:\vspace{-0.5\baselineskip}
\begin{multicols}{2}
\setlength{\columnseprule}{1pt}\noindent
\begin{enumerate}
\renewcommand{\labelenumi}{$\bullet$}
\item $i$ is the imaginary unit.
\item $\mathbb{N} := \{n\in\mathbb{Z}\,|\,n\geq 1\}$
\item $\mathbb{N}_0 := \{0\}\cup \mathbb{N}$
\item For $N\in\mathbb{N}$, $[N] := \{n\in\mathbb{N}\,|\,1\leq n\leq N\}$.
\item For $N\in\mathbb{N}$, $[N]_0 := \{n\in\mathbb{N}_0\,|\,0\leq n\leq N\}$.
\item For $a\in\mathbb{R}$, ${\rm sgn}(a)$ is the sign of $a$:\vspace{-0.3\baselineskip}
\begin{align*}
{\rm sgn}(a) := \left\{\begin{array}{rl}
-1 & (a<0)\\
0  & (a=0)\\
1 & (a>1)
\end{array}\right. .
\end{align*}
\break
\item For $a\in\mathbb{R}$, $\delta_{a}: \mathbb{R}\to\{0,\,1\}$ is \revise{the Dirac delta}:\vspace{-0.5\baselineskip}
\begin{align*}
\delta_{a}(x) = \left\{\begin{array}{ll}
1 & (x=a)\\
0 & (x\neq a)
\end{array}\right..
\end{align*}
\quad\vspace{-2\baselineskip}
\item For a matrix $A$, $A^\trp$ and $A^*$ are the transpose and the conjugate transpose of $A$, respectively.\vspace{-0.5\baselineskip}
\item The symbols ``$\bra{\,\cdot\,}$" and ``$\ket{\,\cdot\,}$'' are often used as description of row vectors and column vectors, respectively. For a column vector $\ket{x}$,  $\bra{x} = \ket{x}^*$. When we describe inner products with them, we can omit a vertical bar: $\bra{x}\ket{y} = \braket{x|y}$.
\end{enumerate}
\end{multicols}
}

\section{Introduction}\label{sec:intro}
In this paper, we examine the concept of a \textit{quantum-walk-replicating random walk}, which we call QWRW. A quantum walk (QW) is often explained as the counterpart of the classical random walk (RW) \cite{K08,Km03,VA12}. However, the properties of a quantum walk are quite different from the latter. Quantum walks were first introduced in the field of quantum information theory \cite{ADZ93,ABNV01}. After that, the characteristic structure of quantum walks was intensively studied by mathematicians, and since then, quantum walks have been an important topic in both fundamental and applied research. Indeed, quantum walks exhibit versatile behavior depending on conditions or settings of time and space, so there are many studies about their mathematical analysis \cite{K02, K05, IKS05, WKKK08, K10, ST12, KLS13, M16}. 
In addition, their unique behavior is useful for implementing quantum structures or quantum analogs of existing models; therefore, the application is considered in fields such as quantum teleportation \cite{YSK21,WSX17}, time series analysis \cite{K19}, topological insulators \cite{AO13, OANK15}, radioactive waste reduction \cite{MIHY11, IMSY15}, and optics \cite{IKMM17}. 
\def\revCZ{Moreover, properties of quantum walks have also been simulated numerically \cite{CZ16},}\revise{\revCZ} or experimentally via the interference of classical light waves \cite{KRS03,JPK04,FIPL06}.

The particularly unique characteristics of a quantum walk are linear spreading and localization. The former means that the deviation of the distribution of a quantum walk is proportional to its run time. The latter implies that there exist specific positions that have non-zero measurement probability after sufficient time evolution. These interesting properties are, however, not easy to intuitively interpret since they are based on the quantum superposition of multiple states. 
On the other hand, modified models of classical random walks are widely considered (e.g., correlated random walk \cite{R81}, L\'{e}vy walk \cite{ZDK15}, Metropolis walk \cite{NOS10}). Such walks are treated as interesting models not only in the field of mathematics but also in natural science \cite{A20}, economics \cite{S06}, informatics \cite{FPRS07}, among others. 
While the distribution of a simple random walk converges to the normal distribution, those of classical random walks, in general, do not necessarily do so, including the ones mentioned above. 
This is why researchers in many fields are interested in them: they can describe more complex transitions of states in real phenomena. 

The relationship between quantum walks and modified classical random walks is an active field of study. A related study is given regarding finite graph structures by Andrade et al \cite{AMF20}. In their study, they show the transition probability matrices of quantum walks as non-homogeneous random walks. 
In the study on $\mathbb{Z}$, the construction of Markov processes is presented by separating the quantum evolution equation into Markovian and interference terms by Romanelli et al \cite{RSSAAD04}. The aim of their decomposition is to show that the linear spreading is derived from coherence. 
That is, if the equation is decoherent, the spread of the probability distribution goes like simple random walks: the standard deviation is $O(\sqrt{t})$, where $t$ is run time \cite{RSAAD04,K07}. In the meantime, Montero studied how to obtain the time- and site-dependent coin operator to generate an intended probability distribution on $\mathbb{Z}$\cite{M17}. Therein, the non-homogeneous random walk on $\mathbb{Z}$ exhibiting the identical probability distribution of quantum walk was discussed as a part of interchanging roles. 

Our study treats the infinite line ($\mathbb{Z}$) and sheds new light on the associated fundamental properties of QWs from the viewpoint of directivity, which is accessible by the notion of QWRWs. 
\def\revCoherent{In conventional quantum walks, the probability of observing a walker in a particular position is calculated via a coherent sum of all probability amplitudes. 
On the other hand, a QWRW provides an individual trajectory to reach a position. 
The probability of observing a walker in a position is calculated via the statistics of individual walkers. 
In other words, the coherent summation of probability amplitudes is transformed into the transition probabilities of walkers at time $n$ and position $x$ on the line in QWRW, which results in the characteristic directivity of walkers.}\revise{\revCoherent} 
While there are previous studies on classical analogs of QWs as mentioned above, we consider that the effects of linear spreading or localization in QW are not examined well 
\def\revPerspective{from such a perspective}\revise{\revPerspective} in the literature.
In particular, concerning the fact that one of the remarkable attributes of QWRW is its ability to track the trajectory of individual walkers, we can obtain insights into the properties of QWs regarding directivity.

More specifically, we characterize the decision of future directions of QWs, which manifests the effect of linear spreading and localization in an unconventional manner. Furthermore, the transition probabilities of the QWRW are spatially and temporally defined and allow to study characteristics of QWs. Moreover, we examine the first return time of the walkers to the origin after departing the origin with respect to RWs and QWs, which is another interesting aspect examined via the notion of QWRWs. We observe that the first return time statistics of QWs are quite different from RWs.

The rest of the paper is organized as follows. Section~\ref{sec:pre} gives the model of one-dimensional { quantum walks and quantum-walk replicating random walks} as preparation. { In Sec. 3 we discuss the directivity of quantum walkers by utilizing QWRWs. Beginning with the discussion on the trajectory of quantum walkers, we discuss the decision of future direction, transition probabilities, and the first return time statistics, which all exhibit different characteristics compared with simple random walks.} Finally, we give a summary and discussion in Sec.~\ref{sec:summary}.

\section{Preliminaries}\label{sec:pre}
In this section, we present quantum-walk-replicating random walk (QWRW). First, we define the two-state quantum walk that QWRW is based on. Second, we describe the way of constructing the QWRW. In this study, we limit the target to $\mathbb{Z}$. We then show that the probability distribution of a QWRW matches that of the corresponding quantum walk by giving an initial condition for both of them.

\subsection{Quantum walk}
First, we give the following unitary matrix: for $x\in\mathbb{Z}$,
\begin{align}
C_x= \twobytwo{a_x}{b_x}{c_x}{d_x},
\end{align}
where $a_x,\ b_x,\ c_x,\ d_x\in\mathbb{C}$ and $a_xb_xc_xd_x\neq 0$. We call $C_x$ a coin of the quantum walk. Here we consider the decomposition of $C$. Let $\ket{\rm L}$ and $\ket{{\rm R}}$ be ${ [1\,\,\,0]^\trp}$ and ${ [0\,\,\,1]^\trp}$, respectively. Using them, we put
\begin{align}
P_x=\ketbra{\rm L}{\rm L}C_x,\quad\quad Q_x=\ketbra{{\rm R}}{{\rm R}}C_x.
\end{align}
Then we obtain
\begin{align}
C_x=P_x+Q_x, \label{decomp}
\end{align}
where $P_x$ and $Q_x$ give the decomposition of $C_x$. $P_x$ and $Q_x$ corresponds to the transition probabilities in the context of simple random walks, respectively.

For a set $(n,\,x)\in \mathbb{N}_0\times\mathbb{Z}$, we define the vector $\varPsi_n(x)\in\mathbb{C}^2$. Here $n$ and $x$ represent the time instant of quantum walks and the position on $\mathbb{Z}$, respectively. Then $\varPsi_n(x)$ stands for the probability amplitude vector of quantum walks on the position $x$ at the time $n$. We define the time evolution of quantum walks as follow:
\begin{align}\label{eq:psiev}
\varPsi_{n+1}(x) = P_{x+1}\varPsi_n(x+1) + Q_{x-1}\varPsi_n(x-1),
\end{align}
which is an analogue of the recurrent formula { of existence probability of a walker} in the context of simple random walks.

Finally, for a set $(n,\,x)$, we define
\begin{align}
\mu_n(x) = \| \varPsi_n(x) \|^2,
\end{align}
where $\mu_n(x)$ now describes the measurement probability of the particle on the position $x$ at the time $n$.

\subsection{Quantum-walk-replicating random walk (QWRW)}
In this subsection, we construct QWRWs. For a set $(n,\,x)\in \mathbb{N}_0\times\mathbb{Z}$ such that $\mu_n(x)>0$, the transition probabilities of the random walks are defined by the following quantity:
\begin{align}\label{pq}
p_n(x) = \fraction{\|P_x\varPsi_n(x)\|^2}{\mu_n(x)},\quad\quad q_n(x) = \fraction{\|Q_x\varPsi_n(x)\|^2}{\mu_n(x)}.
\end{align}
Here for $p_n(x)$ and $q_n(x)$, the following Propositions~\ref{prop:01} and \ref{prop:outflow} hold, which are important to construct the distribution.
\begin{proposition}\label{prop:01}
For a set $(n,\,x)\in \mathbb{N}_0\times\mathbb{Z}$ such that $\mu_n(x)>0$, 
\begin{align}
0\leq p_n(x)  \leq 1\quad\text{and}\quad 0\leq q_n(x) \leq 1.
\end{align}
\end{proposition}

\begin{proof}[\bf Proof]
Here we show this proposition for $p_n(x)$. We can obtain the conclusion for $q_n(x)$ by making a similar discussion.

By the property of norms and the assumption $\mu_n(x)>0$, the inequality 
\begin{align}\label{0p}
0\leq p_n(x)
\end{align}
is trivial. We define $\bra{P_{\rm L}} = [a_x,\, b_x]$, which leads to $\|P\varPsi_n(x)\|^2 = |\braket{P_{\rm L}|\varPsi_n(x)}|^2$. By Cauthy-Schwarz inequality,
\begin{align}
|\braket{P_{\rm L}|\varPsi_n(x)}|^2 \leq \|P_{\rm L}\|^2\cdot\|\varPsi_n(x)\|^2 = (|a_x|^2 +|b_x|^2)\mu_n(x).
\end{align}
Here the relationship $\|P_{\rm L}\|^2 = |a_x|^2 + |b_x|^2 = 1$ holds by the unitarity of the coin $C$. Therefore we obtain the inequality
\begin{align}\label{p1}
\|P\varPsi_n(x)\|^2 \leq \mu_n(x) \,\,\Longleftrightarrow\,\, p_n(x) = \fraction{\|P\varPsi_n(x)\|^2}{\mu_n(x)}\leq 1.
\end{align}
Combining (\ref{0p}) and (\ref{p1}), we obtain the desired result.
\end{proof}

\begin{proposition}\label{prop:outflow}
For a set $(n,\,x)\in \mathbb{N}_0\times\mathbb{Z}$ such that $\mu_n(x)>0$, 
\begin{align}
p_n(x) + q_n(x) = 1.
\end{align}
\end{proposition}

\begin{proof}[\bf Proof]
By unitarity of the coin $C$ and Eq. (\ref{decomp}), we obtain
\begin{align}
\mu_{n}(x) = \|\varPsi_{n}(x)\|^2 = \| C_x\varPsi_n(x) \|^2 = \|P_x\varPsi_n(x) +Q_x\varPsi_n(x)\|^2.
\end{align}
Here the relational expression $\braket{{\rm L|R}} =\braket{{\rm R|L}} =0$ holds, which leads to
\begin{align}
&\|P_x\varPsi_n(x)+Q_x\varPsi_n(x)\|^2\nonumber \\
=\hspace{1.5mm} &\Bigl(\braket{\varPsi_n(x)|C_x^{*}|{\rm L}}\hspace{-1mm}\bra{\rm L}+\braket{\varPsi_n(x)|C_x^{*}|{\rm R}}\hspace{-1mm}\bra{{\rm R}}\Bigr)\Bigl(\ket{\rm L}\hspace{-1mm}\braket{{\rm L}|C_x|\varPsi_n(x)}+\ket{{\rm R}}\hspace{-1mm}\braket{{\rm R}|C_x|\varPsi_n(x)}\Bigr)\nonumber \\
 =\hspace{1.5mm} & \|P_x\varPsi_n(x)\|^2 + \|Q_x\varPsi_n(x)\|^2.
\end{align}
Therefore,  
\begin{align}
\|P_x\varPsi_n(x)\|^2 + \|Q_x\varPsi_n(x)\|^2 = \mu_n(x) \,\,\Longleftrightarrow\,\, p_n(x) +q_n(x) = 1.
\end{align}
This is the desired equation.
\end{proof}
By the propositions above, we can define QWRW as follows:

\begin{defin.}{\bf [quantum-walk-replicating random walk (QWRW)]}\\
Let $\{\varPsi_n\}_{n\in\mathbb{N}_0}$ be the quantum walk defined by (\ref{eq:psiev}), i.e., 
\begin{align}
    \varPsi_{n+1}(x) = P_{x+1}\varPsi_n(x+1) +Q_{x-1}\varPsi_n(x-1),\quad \varPsi_0(x) = \delta_0(x)\varphi_0
\end{align}
with $\|\varphi_0\| =1$. The quantum-walk-replicating random walk (QWRW) $\{S_n\}_{n\in\mathbb{N}_0}$ satisfies the following evolution: for $(n,\,x)\in\mathbb{N}_0\times\mathbb{Z}$ such that $\mu_n(x)>0$,
\begin{align}
    \mathbb{P}(S_{n+1} =x+\xi\,|\,S_n=x) = \left\{
    \begin{array}{ll}
    p_n(x) & :\xi=-1\\
    q_n(x) & :\xi=+1\\
    0 & :\text{otherwise}
    \end{array}\right.,
\end{align}
where $p_n(x) = \fraction{\|P_x\varPsi_n(x)\|^2}{\mu_n(x)}$ and $q_n(x) = \fraction{\|Q_x\varPsi_n(x)\|^2}{\mu_n(x)}$.
\end{defin.}

Let $\nu_n(x)$ be the probability that a particle following QWRW (a QWRWer) exists on $x\in\mathbb{Z}$ at the time $n\in\mathbb{N}_0$. In other words, $\nu_n(x)$ is defined by the initial distribution and the following recurrent formula:
\begin{align}\label{tnuev}
\nu_{n+1}(x) = p_n(x+1)\nu_n(x+1) + q_n(x-1)\nu_n(x-1).
\end{align}
Moreover, we put
\begin{align}
\nu_n = [\cdots,\,\nu_n(-1),\,\nu_n(0),\,\nu_n(1),\,\cdots]^\trp
\end{align}
and call it the distribution of QWRWs at time $n$.

Incidentally, we have the following lemma:
\begin{lemma}\label{lem:mu}
For $n\in\mathbb{N}_0$,
\begin{align}
\mu_{n+1}(x) = p_n(x+1)\mu_n(x+1) + q_n(x-1)\mu_n(x-1).
\end{align}
\end{lemma}
\begin{proof}[\bf Proof]
By Eq. (\ref{eq:psiev}), 
\begin{align}
\mu_{n+1}(x) &= \|\varPsi_{n+1}(x)\|^2 = \|P_{x+1}\varPsi_n(x+1)+Q_{x-1}\varPsi_n(x-1)\|^2 \nonumber\\
&= \scalebox{0.9}[1]{$\Bigl(\braket{\varPsi_n(x+1)|C_{x+1}^{*}|{\rm L}}\hspace{-1mm}\bra{\rm L}+\braket{\varPsi_n(x-1)|C_{x-1}^{*}|{\rm R}}\hspace{-1mm}\bra{{\rm R}}\Bigr)\Bigl(\ket{\rm L}\hspace{-1mm}\braket{{\rm L}|C_{x+1}|\varPsi_n(x+1)}+\ket{{\rm R}}\hspace{-1mm}\braket{{\rm R}|C_{x-1}|\varPsi_n(x-1)}\Bigr)$} \nonumber\\
&= \|\braket{{\rm L}|C_{x+1}|\varPsi_n(x+1)}\|^2 + \|\braket{{\rm R}|C_{x-1}|\varPsi_n(x-1)}\|^2 = \|P_{x+1}\varPsi_n(x+1)\|^2 + \|Q_{x-1}\varPsi_n(x-1)\|^2.
\end{align}
By the definition of $p_n(x)$ and $q_n(x)$ (\ref{pq}), we obtain the desired conclusion.
\end{proof}
Using Lemma \ref{lem:mu}, we obtain the following fact: if and only if we assume $\nu_0 = \mu_0$, the distribution of QWRWs and QWs coincident completely. Therefore, we can draw the following theorem:
\begin{theorem}\label{thm:match}
\begin{align}
\nu_0 = \mu_0 \,\,\Longleftrightarrow\,\, \nu_n = \mu_n\,\,\text{for all}\,\,n\in\mathbb{N}_0.
\end{align}
\begin{proof}[\bf Proof]
We assume that for time instant $n\in\mathbb{N}_0$
\begin{align}
\nu_n = \mu_n.
\end{align}
Then using the relational expression (\ref{tnuev}) and the assumption above, we have
\begin{align}
\nu_{n+1}(x) = p_n(x+1)\mu_n(x+1) +q_n(x-1)\mu_n(x-1)
\end{align}
for any $x\in\mathbb{Z}$. Therefore, by the Lemma \ref{lem:mu}, we obtain
\begin{align}
\nu_{n+1}(x) = \mu_{n+1}(x).
\end{align}
Reminding that this holds for any $x\in\mathbb{Z}$, we obtain the desired conclusion.
\end{proof}
\end{theorem}

Here we compare our theory with previous studies. Ours extends Theorem~1 in \cite{AMF20} to an infinite graph if we consider numbers of position $x\in\mathbb{Z}$ as labels of vertices on the infinite graph $G(\mathbb{Z},\,E_{\mathbb{Z}})$ with $E_{\mathbb{Z}}=\{ (x,\,y)\in\mathbb{Z}^2\,|\, |x-y| =1\}$. Our definition of transition probabilities (\ref{pq}) corresponds to the case that $\upsilon(u,\,t)>0$ and $(u,\,v)\in E$ in Eq.~(18) in their results, which is the case that $\mu_n(x)>0$ and $(x,\,y)\in E_{\mathbb{Z}}$ in our contexts. Note that $\rho(v,\,c,\,t+1)$ in their paper corresponds to the left (resp. right) chirality of $\varPsi_{n+1}(x-1)$ ($\varPsi_{n+1}(x+1)$), whose value matches the left (right) component of $P_x\varPsi_n(x)$ ($Q_x\varPsi_n(x)$). The difference of our study from their result is that the identification between the distribution of quantum walk ($\mu_n$) and QWRW ($\nu_n$) is the theorem as we showed in Theorem~\ref{thm:match}  while they assume the match between quantum walk ($\upsilon(v,\,t)$) and non-homogeneous random walk ($\pi_v(t)$).

In \cite{M17}, almost the same model is generated as a result of interchanging roles, which solve the problem of how a time- and site-dependent random walk mimics the properties of a quantum walk or the opposite situation. There time- and site-dependent random walk and quantum walk are combined by equating both of the net fluxes, which are
\begin{align}
J_n^{\rm (RW)}(x) = (q_n(x)-p_n(x))\nu_n(x)\quad\text{and}\quad J_n^{\rm (QW)}(x) = \|Q\varPsi_n(x)\|^2-\|P\varPsi_n(x)\|^2
\end{align}
respectively, in our contexts. However, this theory also needs the assumption that the distribution of time- and site-dependent random walk and that of quantum walk are the same, which are described as $\rho(n,\,t)$ in \cite{M17}. In this sense, our introduction of the distribution is made in the opposite direction to their paper.

\subsection{Demonstrations}
\begin{figure}[p]
\centering
\includegraphics[width=0.8\linewidth]{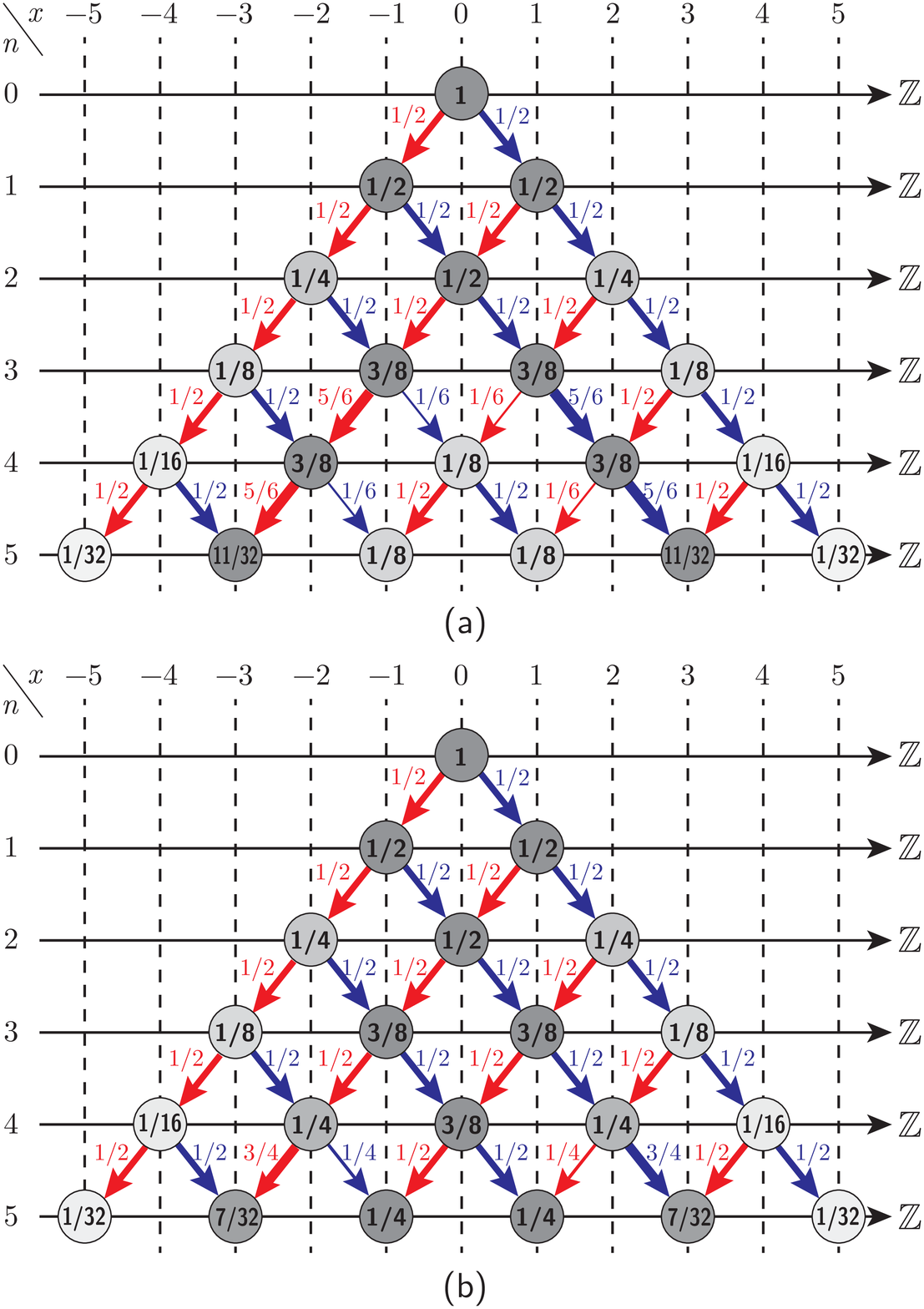}
\caption{
Transition of the distribution $\nu_n$ of the QWRW corresponding to \revise{(a) Ex. 1 and (b) Ex. 2}. Red and blue arrows represent $p_n(x)$ and $q_n(x)$ from each set $(n,\,x)$, respectively. Values on each coordinate $(n,\,x)$ represents the corresponding value of $\nu(x)$. For the coordinates that don't have any display about $\nu_n(x)$, their values are $0$.
}
\label{fig:qpachi}
\end{figure}

\begin{figure}[t!]
\centering
\includegraphics[width=170mm]{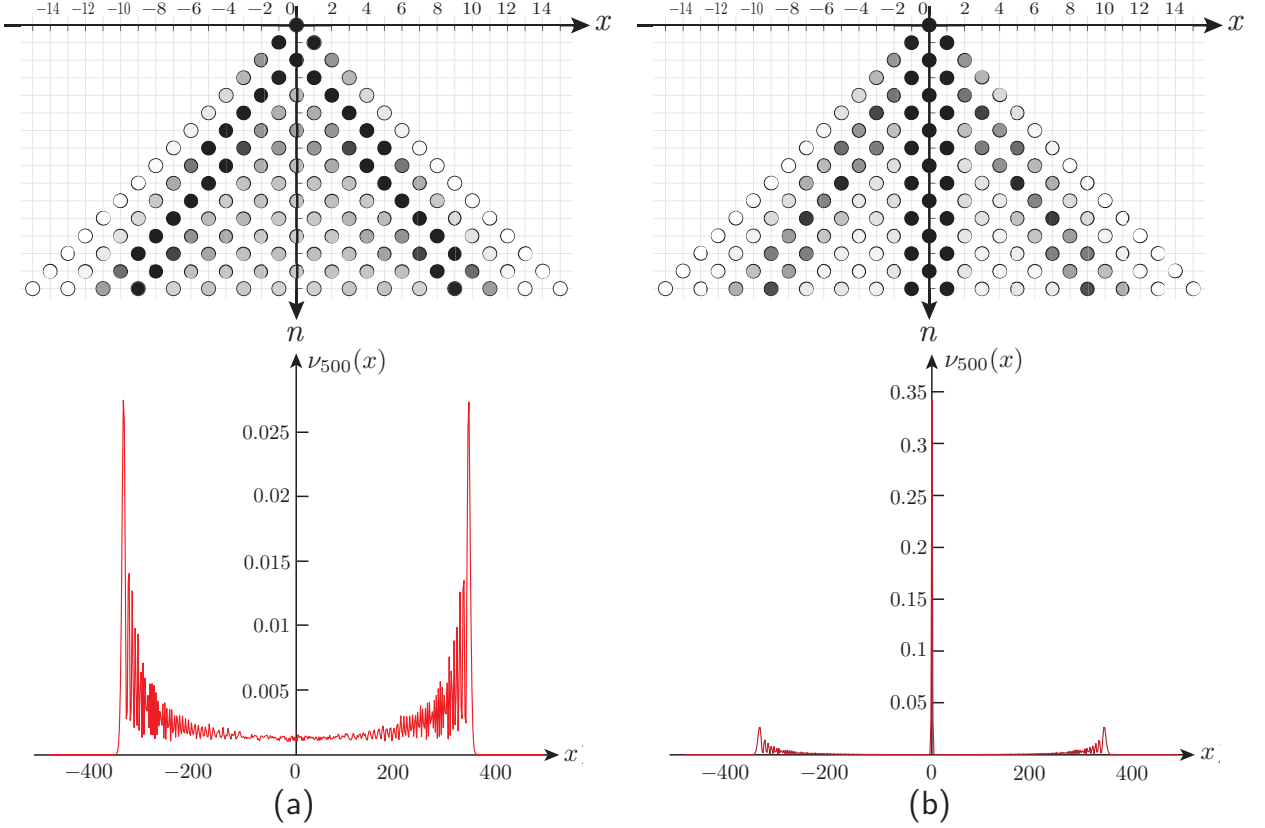}
\caption{Graphical expressions of QWRW for {\sf (a)} example 1 (site-homogeneous model) and {\sf (b)} example 2 (one-defect model). The upper figure of each panel illustrates the space-time diagrams of the probability distribution. Circular markers are located at the coordinates $(n,\,x)$ such that $\nu_n(x)>0$ with their color indicating the value of $\nu_n(x)$. The darker the color is, the larger the value of $\nu_n(x)$ is. 
The lower figure of each panel shows probability distribution of QWRW $\nu_n(x)$ at the time instant $n=500$, which exactly matching the distribution of the original QWs. 
\newline{}\vrule height 0.3mm width 165mm}
\label{fig:qwrwex}
\end{figure}

Here we show two examples, which are treated in the numerical studies later.\\

\noindent{\bf Example 1. [Site-homogeneous model]}\\
We set
\begin{align*}
C_x = \fraction{1}{\sqrt{2}}\twobytwo{1}{1}{1}{-1} \,\,\text{for any $x$,\quad and}\quad \varPsi_0(0) = \frac{1}{\sqrt{2}}\onebytwo{1}{i}.
\end{align*}

\def\revA{In this case, transition and existence probabilities are calculated like Fig.~\ref{fig:qpachi}(a). These calculations are detailed in Appendix~A.1. The probability remains the same as the case of a symmetric simple random walk until time $n=3$. 
However, the distribution starts to differ at time $n=4$.
From there on, the effects of linear spreading become apparent. The resulting probability distribution $\nu_n(x)$ appears as Fig.~\ref{fig:qwrwex}(a). This graph shows the peaks on the ends; these correspond to linear spreading. Localization does not appear in this case.}\revise{\revA}

The quantum walk model based on this example is well known as one of the most fundamental models of quantum walks and is introduced as {\it symmetric model} in Konno \cite{K05}. The graph of the probability function also matches that of Fig. 3 in Konno \cite{K05}.\\

\noindent{\bf Example 2. [One-defect model]}\\
We set
\begin{align*}
C_x = \left\{\begin{array}{ll}
\twobytwo{1}{0}{0}{1} & (x=0) \vrule width 0pt height 9pt depth 13pt\\
\fraction{1}{\sqrt{2}}\twobytwo{1}{1}{1}{-1} & (x\not =0)
\end{array}\right. \!\text{,\quad and}\quad \varPsi_0(0) = \frac{1}{\sqrt{2}}\onebytwo{1}{i}.
\end{align*}

\def\revB{In this case, transition and existence probabilities are calculated like Fig.~\ref{fig:qpachi}(b). These calculations are detailed in Appendix~A.2. The probability is the same as the case of symmetric simple random walks until $n=4$. 
However, at time $n=5$ QWRWers are likely to go outside on $x=\pm 1$, which results in a different probability distribution from the symmetric simple random walk.  The resulting probability distribution $\nu_n(x)$ is shown in Fig.~\ref{fig:qwrwex}(b). While this example has peaks on the ends similar to Ex. 1, it also has an extremely high peak on the origin, as this example shows both linear spreading and localization.}\revise{\revB}

The quantum walk model based on this example is often called {\it one-defect quantum walk}, which exhibits localization at $x=0$ as well as linear spreading.

\section{Directivity of QWRWers}\label{sec:main}
In this section, we introduce graphical representations of QWRW as our main results. Comparing them with those of simple random walk, whose walkers go to both sides with probability $1/2$ regardless of time and site, we point out how linear spreading and localization play roles as walk controllers. 

\subsection{Trajectories}
One of the significant benefits of QWRWs is that the notion of the path of a walker is valid because the walker follows a classical probability at each position. A real quantum walk is not a stochastic process, so it is not permitted to observe its path in such a classical way. In this regard, QWRW plays an important role as the visualizer of quantum walks. Here we demonstrate the path trajectories of the walkers of QWRWs. 

\begin{figure}[t!]
\begin{center}
\includegraphics[width=\textwidth]{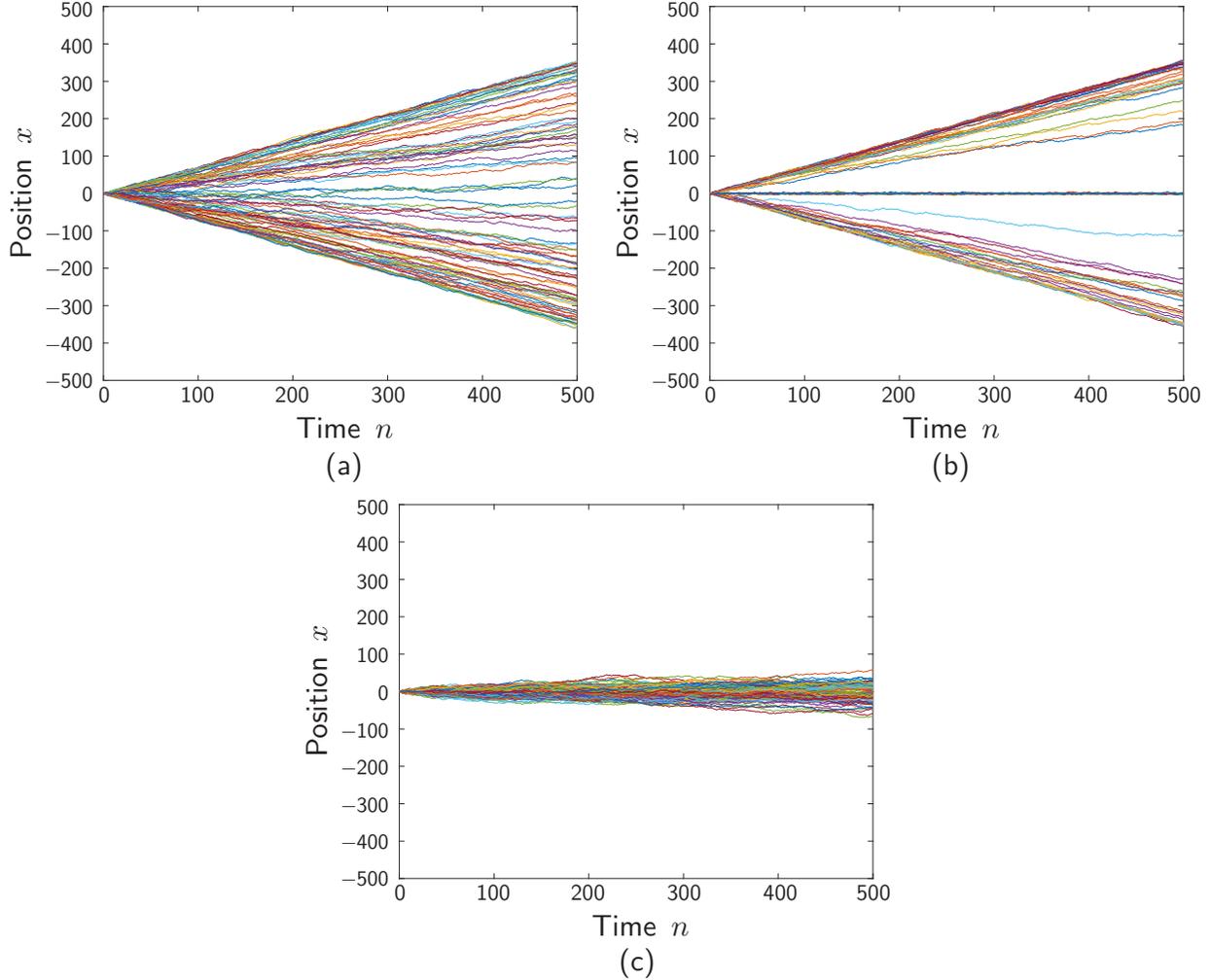}
\caption{The paths of individual walkers in QWRWs for final time $N=500$ and $100$ walkers for {\sf (a)} example 1 (site-homogeneous model), {\sf (b)} example 2 (one-defect model), and {\sf (c)} simple random walk. The linear spreading of quantum walks is visible as quasi-ballistic trajectories towards the edges. Paths staying around the origin in Ex. 2 (b)  localization by the defect coin in the quantum walk.
\newline{}\vrule height 0.3mm width 165mm}
\label{fig:path}
\end{center}
\end{figure}

The curves in Figs. \ref{fig:path}(a) and (b) show the path trajectories of the walkers of QWRWs, which correspond to Exs. 1 and 2, respectively. For comparison, we also show the path trajectories of simple random walk in Fig. \ref{fig:path}(c). The horizontal and vertical axis respectively denote time and position. The number of walkers shown therein is $100$.

There are many paths that spread towards the end sections resulting in linear spreading. While in genuine QW, linear spreading and localization are the effects of many interfering wave amplitudes, in QWRWs this effect can be observed as a trace of an individual walker. 

In Fig.~\ref{fig:path}(a), the corresponding QWRW has the property of linear spreading, but does not exhibit localization. Linear spreading carries walkers radially, and the range of terminus of walkers is much wider than that of simple random walkers. Moreover, the farther from the origin the position is, the higher the density of walkers, which corresponds to the highest peaks of the probability distribution shown in Fig.~\ref{fig:qwrwex}.

On the other hand, in Fig.~\ref{fig:path}(b), both linear spreading and localization work on walkers. As with Ex. 1, some walkers leave away from the origin. However, many walkers wander around the origin, which is a clear difference from the previous example. Furthermore, comparing the area $-200<x<200$ with Fig.~\ref{fig:path}(a), the center area except for the origin is almost free of walkers. This range is also much narrower than the running range of the simple random walker.  These different observations from Ex. 1 are affected by localization. This property catches walkers and prevents them from leaving away. Meanwhile, the walkers that are not captured by localization go far under the influence of linear spreading. Eventually, walkers are divided between far travelers and stayers on the origin.

\subsection{Future direction}\label{ssec:fd}

 \begin{figure}[t!]
\begin{center}
\includegraphics[width=\textwidth]{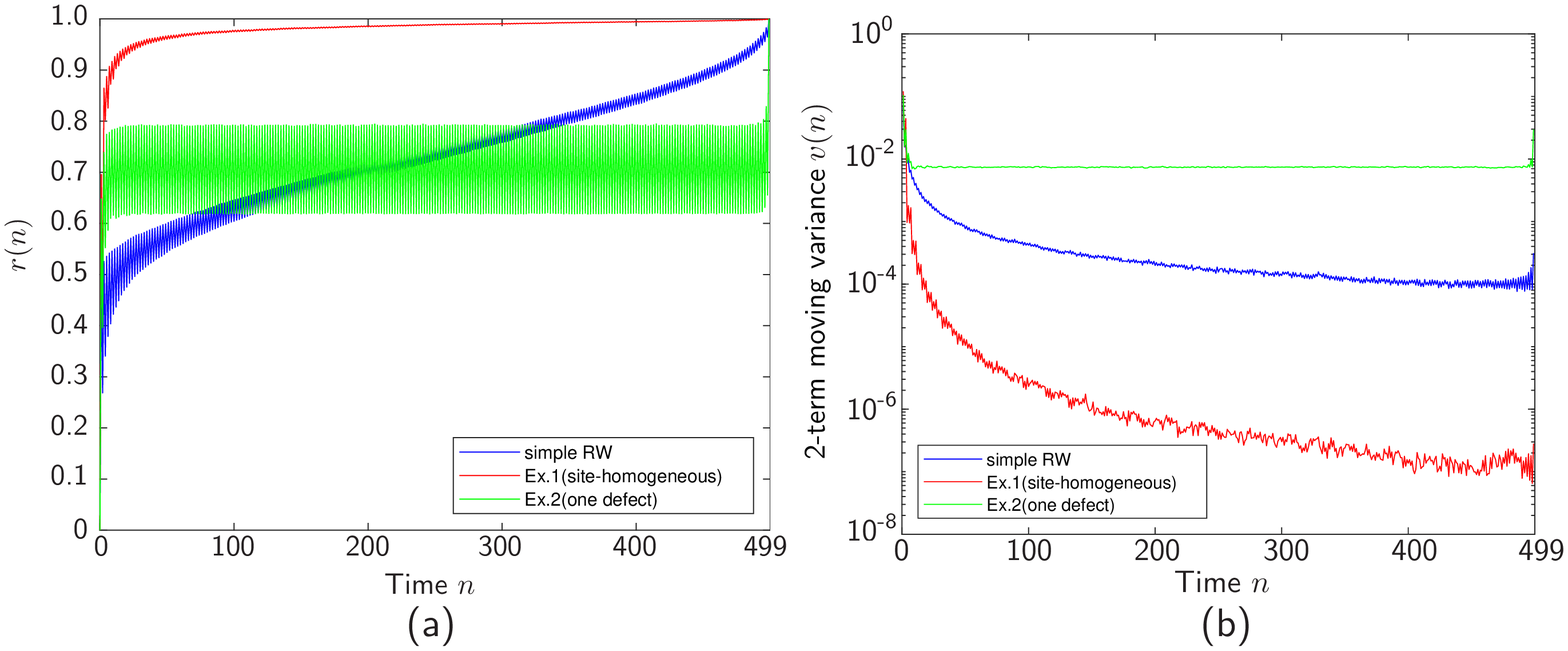}
\caption{
{\sf (a)} The degree of deciding future direction is characterized through the analysis of paths derived via QWRWs or the simple random walk ($N=499$, $K=100000$). According to Eq. (\ref{eq:rn}), walkers are counted to the numerator of $r(n)$ in the only case that the sign at time $n$ matches the final one. Note that it is not counted in case of $S_n = 0$. QWRWs of Exs. 1 and 2 exhibit a dramatic increase right after the initial time, which is another manifestation of linear spreading.
{\sf (b)} The 2-term moving variance (MV) $v(n)$ of $r(n)$. QWRW of Ex. 2 exhibits a larger variance, which clearly indicates localization.
 \newline{}\vrule height 0.3mm width 165mm
}
\label{fig:future}
\end{center}
\end{figure}

We further examine the properties made observable by QWRWs through the analysis of individual trajectories. Here, the final time instant is given by $N\in\mathbb{N}$. Let $K$ be the number of walkers of QWRWs or the simple random walk and $S_n^{(k)}$ as the position of the $k$-th walkers at time $n$ with $k\in[K]$ and $n\in[N]_0$. Moreover, using them, we define $r(n)$ as the following to characterize the relevance of the current (time: $n$) position and the future (time: $N$) position: 
\begin{align}\label{eq:rn}
r(n) = \fraction{1}{K}\times \#\{k\in[K]\,|\,{\rm sgn}(S_N^{(k)}){\rm sgn}(S_n^{(k)})=1\}.
\end{align}
Here ${\rm sgn}(S_N^{(k)}){\rm sgn}(S_n^{(k)})=1$ means that the $k$-th walker is on the same side as the final position with respect to the origin at time $n$. On the other hand, ${\rm sgn}(S_N^{(k)}){\rm sgn}(S_n^{(k)})=-1$ means that the $k$-th walker is on the opposite side of the final position with respect to the origin at time $n$. In this way, ${\rm sgn}(S_N^{(k)}){\rm sgn}(S_n^{(k)})$ is the benchmark to investigate when walkers effectively determine the directions they arrive in the future. If a walker maintains ${\rm sgn}(S_N^{(k)}){\rm sgn}(S_n^{(k)})=1$, we can interpret the walker determines its own evolving direction. Based on that, we can consider that $r(n)$ is a figure-of-merit to investigate the ratio of walkers that determines their future direction.

The red and green curves in Fig. \ref{fig:future}(a) show $r(n)$ regarding QWRWs of Exs. 1 and 2, respectively. In addition, the blue curve in Fig. \ref{fig:future}(a) indicates $r(n)$ for the simple random walk. 

We can clearly observe that $r(n)$ of QWRWs increases dramatically soon after the time evolution begins after $n = 0$, i.e., the future direction of a walker is highly determined by the position in its early stage. This observation is affected by linear spreading; this property pulls walkers from the origin and prevents returning there again. On the other hand, the jump-up of $r(n)$ of the one-defect QWRW (Ex. 2) is not higher than that of the no-defect QWRWs (Ex. 1) and maintains the value within a certain range. In addition, even after $r(n)$ of Ex. 2 is reached and overtaken by that of the simple random walk, the stay of value continues, and at the very end, goes to 1. This is the effect that localization makes some walkers wander around the origin, which repeats crossing $x=0$.

Another observable feature is oscillation. The ratio $r(n)$ gets higher and lower value when $n$ is odd and even, respectively, which causes oscillation for all the curves. This is because some walkers are on the origin at even time instance because they are not counted on the numerator of $r(n)$ (see Eq.~(\ref{eq:rn})). In response to this fact, we set the final time instance of this simulation as an odd number so as to have $r(n)$ reach $1$ in the end. The amplitude of oscillation can be estimated by the 2-term moving variance $v(n)$ of $r(n)$  by the following equation, see Appendix A for details.
\begin{align}\label{eq:mv}
v(n) = \fraction{1}{4}(r(n)^2-2r(n)r(n-1)+5r(n-1)^2).
\end{align}
The results are shown in Fig. \ref{fig:future}(b). We observe that $v(n)$ decreases dramatically right after the jump-up in QWRWs of Ex. 1. This is because most of the walkers  leave from the origin right away with linear spreading and never return there again. Conversely, in QWRW of Ex. 2, $v(n)$ does not exhibit a drastic decrease, which stays at a constant value of approximately $10^{-2}$. This means that many walkers of the one-defect QWRW wander around the origin right before the final time; that is the effect of the other characteristics of quantum walks: localization.

\subsection{Transition probability}

\begin{figure}[t!]
\centering
\includegraphics[width=\textwidth]{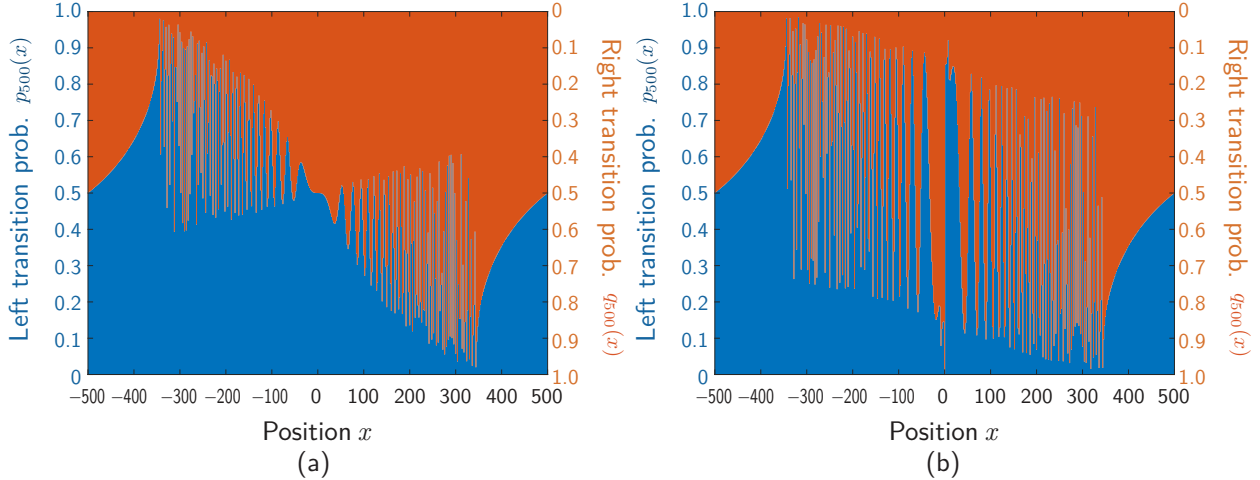}
\caption{
The transition probabilities $p_{500}(x)$ and $q_{500}(x)$ for QWRWs generated from QWs for {\sf (a)} Ex. 1 (site-homogeneous) and {\sf (b)} Ex. 2 (one-defect). It holds that { $p_{500}(x) + q_{500}(x) = 1$}, so left ($p_{500}(x)$, blue) and right ($q_{500}(x)$, orange) transition probabilities are expressed as a stacked bar graph.
\newline{}\vrule height 0.3mm width 165mm}
\label{fig:transprob}
\end{figure}

We examine the expressions of transition probabilities $p_n(x)$ and $q_n(x)$. 
The sum of $p_n(x)$ and $q_n(x)$ is equal to 1; hence our interest is in  the imbalances between $p_n(x)$ and $q_n(x)$. The red and red portion of the color bars in Fig.~\ref{fig:transprob} indicates the amount of $p_n(x)$ and $q_n(x)$, respectively. 

A common attribute observed in Figs.~\ref{fig:transprob}{ (a) and (b)} is that there are peaks of $p_n(x)$ and $q_n(x)$ on the left and right sides, respectively. For $x\leq -350$ and $x\geq 350$ all examples have identical transition probablities, following a simple quasi-continuous function of $x$. This part of the transition probabilities indicates that individual walkers that go far away from the origin tend to go farther, which is related to the linear spreading property of QWs. 

In the center region of Fig.~\ref{fig:transprob}, the transition probabilities are fluctuating quickly as a function of $x$. These fluctuations are { calm} around the origin, but { are violent outside $x=\pm 50$}. The exact position and height of these maxima and minima in $p_n(x)$ and $q_n(x)$ also changes with $n$. The cause of these fluctuations are not clarified, but we hypothesize that they are the classical representation of self-interference of the quantum walkers.

A { significant} difference between two examples shown in Fig.~\ref{fig:transprob} is the distribution around the origin. 
In Fig.~\ref{fig:transprob}(a), $p_n(x)$ exhibits a peak on the left side of the origin while $q_n(x)$ shows a peak on the right side of the origin. This means that the QWRWers starting from the origin equally likely go to the left side or to the right side. This is strongly related to the fact that the probability distribution $\nu_n(x)$ in Ex. 1 exhibits its highest peaks on both left and right sides (Fig.~\ref{fig:qwrwex}(a)). 

On the other hand, in Fig.~\ref{fig:transprob}(b), $p_n(x)$ and $q_n(x)$ show the local maximum on the right and left sides, respectively, meaning that the QWRWers are guided toward opposite directions alternatively around the origin. Namely, QWRWers are highly likely locked in around the origin, which is a manifestation of localization (Fig.~\ref{fig:qwrwex}(b)). This is considered to be a result of the defect located near the origin. Furthermore, from the transition probabilities calculated via QWRWs, we can foresee certain underlying smooth structures plus highly oscillatory ones in Figs.~\ref{fig:transprob}(a) and (b), which is one of the interesting future study.

\subsection{First return time}

\begin{figure}[t!]
\begin{center}
\includegraphics[width=\textwidth]{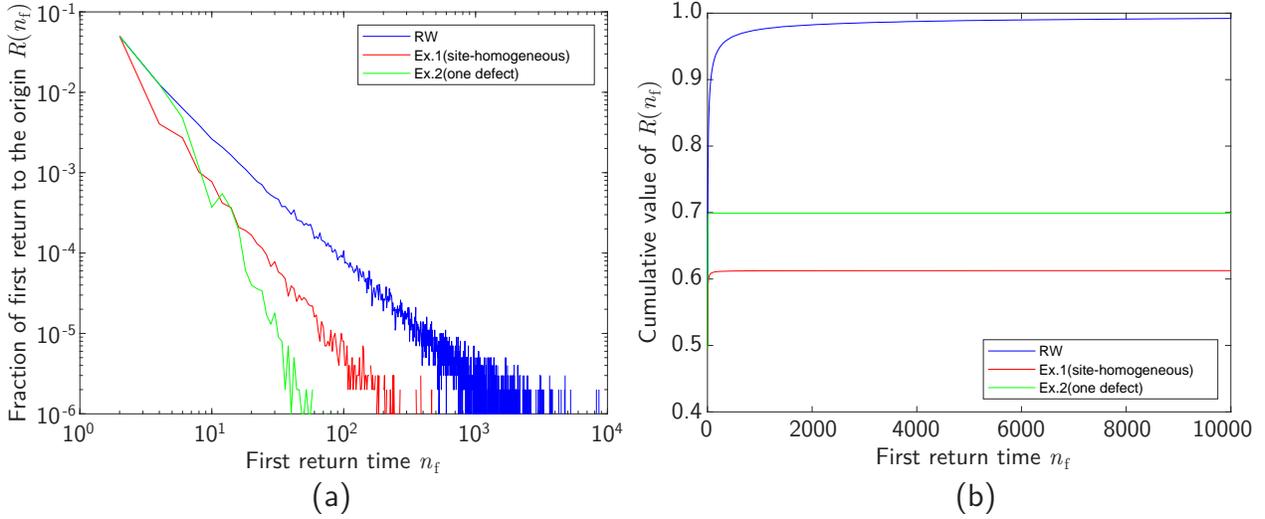}
\caption{{
{\sf (a)} The relationship between time $n_{\rm f}$ and the ratio of walkers whose first return time is the time, which is described as $R(n_{\rm f})$ ($N=10^4$, $K=100000$). 
{\sf (b)} Cumulative value of $R(n_{\rm f})$, whose setting is the same as (a). In case of simple random walk, site-homogeneous QWRW, and one-defect QWRW, $R(10000)$ is approximately $0.992$, $0.613$, and $0.699$, respectively.
}
\newline{}\vrule height 0.3mm width 165mm}
\label{fig:frt}
\end{center}
\end{figure}

The first return time concerns whether a walker departing from the origin comes back to the origin and if so, the first return time is the time when the walker comes back to the origin first, which is denoted by $n_{\rm f}$. This is one of the interesting figures to characterize the directivity of walkers. It is well known that the probability of a walker to be back to the origin at time $n_{\rm f}$ for a symmetric one-dimensional random walk follows power law $n_{\rm f}^{-3/2}$. Furthermore, the cumulative return probability is 1, meaning that the random walker surely comes back to the origin at some moment.

Fig. \ref{fig:frt}(a) summarizes the fraction of walkers whose first return time is time $n_{\rm f}$, which is denoted by $R(n_{\rm f})$. The blue, red, and green curves represent the result with respect to simple RW, site-homogeneous QWRW, and one-defect QWRW, respectively. Here we run in a total of 100,000 random walkers. The simple RW (red curve) in Fig. \ref{fig:frt}(a) follows the abovementioned $n_{\rm f}^{-3/2}$ trend. In addition, Fig. \ref{fig:frt}(b) characterizes the cumulative value of $R(n_{\rm f})$ as a function of $n_{\rm f}$. It should be noted that the red curve by the simple RW is approaching unity, manifesting that the return probability of simple RW is one.

In contrast, the site-homogeneous QWRW exhibits smaller $R(n_{\rm f})$ values for any $n_{\rm f}$ than simple RW in Fig. \ref{fig:frt}(a), which indicates that the walker is less likely to come back to the origin at time $n_{\rm f}$ than for a simple RW. Indeed, in Fig. \ref{fig:frt}(b), the cumulative value of $R(n_{\rm f})$ undergoes a plateau, indicating strongly that some of the walkers never return to the origin. This is another manifestation of the linear spreading of QW in the format of the first return time discussion. Indeed, the red curve in Fig. \ref{fig:frt}(a) approximately numerically follows power-law statistics with $n_{\rm f}^{-2}$. A strict derivation of this distribution could give further insights into the behavior of the QWRWs, but so far this remains an open challenge.

In the one-defect model (green curve), $R(n_{\rm f})$ shows the same order of value as that of simple RW until $n_{\rm f}=8$, but after that, it drastically decreases. This indicates that the defect coin strongly pulls some walkers back to the origin by localization to the origin (until $n_{\rm f}=8$), but such a character quickly decays, and most of the walkers run away because of linear spreading. Similar to the former site-homogeneous case, the cumulative value of $R(n_{\rm f})$ sees a plateau, indicating that some of the walkers spread away from the origin. At the same time, it should be noted that the cumulative value of $R(n_{\rm f})$ by one-defect QW is larger than that by site-homogeneous QW, indicating that the localization by the defect contributes to the increase of the returns of the walkers.

Connecting the results of the future direction (Fig. 4) and first return time (Fig. 6), we can also see that approximately 38\% of walkers for site-homogeneous QWRW never return to the origin, which means that their future direction must stay on the same side after time $n=1$. Indeed, compared with future directions shown in Fig.~\ref{fig:future}(a), $r(n)$ is larger than about 0.38. Similarly, for all the models of walk and time instance $n$, $r(n)$ is larger than the cumulative value of $R(n)$, and the difference is related to the walkers that cross the origin several times because of wandering.

\section{Summary and Discussion}\label{sec:summary}
In this study, we examine the directivity of quantum walks via its random walk replica, which is called quantum-walk-replicating random walk (QWRW). Although the theoretical approach to deal with quantum walks in classical random walks (that is, QWRW) has been studied in the literature, the insights and viewpoints made possible by QWRW are not well characterized, especially the directivity of quantum walks. We start with the derivation of trajectories of QWRW, followed by the demonstration of transition probabilities to left and right, which coincides with the linear spreading and localization properties of QW. Furthermore, we demonstrate that the future direction of QWRW is determined at the very early stage of the time evolution, which is another representation of the linear spreading attribute. Finally, the first return time of QWRW is examined where the scaling properties are different from the simple RW. With the site-homogeneous coin and one-defect coin examples, we can clearly observe the linear spreading and localization effects in the first return time statistics.

For future studies, more avenues of research remain. A deeper mathematical understanding of QWRWs is desirable and several open questions remain. For example: what are the asymptotic behavior of transition probabilities $\lim_{n\rightarrow\infty}p_n(x)$ and $\lim_{n\rightarrow\infty}q_n(x)$ for each $x\in\mathbb{Z}$? Can the metrics to describe the future directions be represented by an analytical formula? Besides, quantum walks strongly depend on initial conditions as studied in \cite{TFMK03}; therefore, the initial state dependency is a promising future direction to QWRW. 

\def\revHermite{So far, the coin operator has also been restricted to be a Hermite matrix. 
Extension to non-Hermitian systems \cite{CWZ17,ZLC20} would significantly enhance the generality of the related discussions.}\revise{\revHermite}

\def\revEntangled{The present study only examines single-particle systems. QWRW approaches for multi-particle systems, including entanglement, could provide even deeper insights into the relationship between classical RWs and QWs. Entangled QWRWs could also potentially have computational advantages over full quantum simulations, although this remains an open question.}\revise{\revEntangled}

\def\revEfficiency{In particular, QWRWs generate walkers exhibiting equivalent statistics as the measurement result of a QW. Computationally, the cost is also much lower under certain assumptions; once the transition probabilities at time $n$ and the position $x$ are known, calculating an individual trajectory of the QWRW is very fast and will be $O(n)$ for trajectories until time $n$. A comparable QW would necessarily need to calculate the fully self-interacting wave function, which requires matrix products and is $O(n^2)$.

However, in our current work, the correct transition probabilities $p_n$ are directly calculated from the wave function, which requires a complete QW simulation initially, as this is how we ensure that the resulting QWRW shares all important properties with the original QW. 
As related thoughts in Andrade et al. \cite{AMF20} outline, there are potential cases where $p_n$ could be obtained directly. 
In those cases, which for now appear to be rare, a simulation of QWRW would overall be more efficient than a QW simulation. 
Whether it is possible to obtain the transition probabilities $p_n$ and $q_n$ in general through some clever formulae that avoid calculating the wave function of the corresponding QW remains an open problem.}\revise{\revEfficiency}

\def\revPachinko{Besides the numerical simulation of QWRWs, the physical implementation of QWRWs is an intriguing issue. In QWRW, the walker transits to the left or right depending on the transition probability at any given time and position. This could be implemented physically, for example in a mechanical system where a ball, the walker, rolling down a plane encounters an array of pins that deflect the ball to left or right in a probabilistic manner, just like in ``pachinko" \cite{pachinko}. The precision of the deflection rates may, however, turn out to be a limiting factor for large QWRW. To its advantage, such a mechanical system also ensures an effectively discrete-time evolution of the ball's pathway, corresponding to the rows in a classical ``pachinko" machine. Ultimately, much faster physical processes may also allow for the construction of quasi-``pachinko" machines, such as electronics or photonics.}\revise{\revPachinko}

\def\revApplication{A QWRW trajectory can also be the starting point of further applications. 
In the literature, a chaotically oscillating time series has been experimentally utilized in efficiently solving multi-armed bandit problems \cite{NMHSOH18}. 
Furthermore, Okada et al. \cite{OYCIHN22} demonstrated a theoretical model that accounts for the acceleration of solving the two-armed bandit problem within the framework of correlated random walks. 
These findings imply that a QWRW may be an exciting resource for solving such bandit problems.}\revise{\revApplication}

\appendix
\def\thesection{Appendix \Alph{section}}
\revise{
\section{Calculations of probabilities on QWRW}
\subsection{Ex. 1 [Site-homogeneous model]}
}\noindent
\def\revCalA{The coin matrix $C_x$ is decomposed as follows: for all $x\in\mathbb{Z}$,
\begin{align}\label{eq:decompose}
P_x= \ketbra{\rm L}{\rm L}C_x = \fraction{1}{\sqrt{2}}\twobytwo{1}{1}{0}{0} =: P,\quad\quad Q_x=\ketbra{\rm {\rm R}}{\rm {\rm R}}C_x=\fraction{1}{\sqrt{2}}\twobytwo{0}{0}{1}{-1} =: Q.
\end{align}
The probability at time $n=0$ is defined as
\begin{align}
\nu_0(0) = \mu_0(0) = 1.    
\end{align}
The probabilities $p_0(0)$ and $q_0(0)$ are calculated as follows:
\begin{align}
    p_0(0) = \fraction{\|P\varPsi_0(0)\|^2}{\mu_0(0)}  = \fraction{1}{2},\,\,\,\,
    q_0(0) = \fraction{\|Q\varPsi_0(0)\|^2}{\mu_0(0)}  = \fraction{1}{2}.
\end{align}
Therefore, the existence probabilities at time $n=1$ are obtained as follows:
\begin{align}
    \nu_1(-1) = p_0(0)\nu_0(0) = \fraction{1}{2},\,\,\,\,
    \nu_1(1) = q_0(0)\nu_0(0) = \fraction{1}{2}
\end{align}
The probabilities $p_1(\pm 1)$, $q_1(\pm 1)$ are calculated as follows:
\begin{align}
    &p_1(-1) = \fraction{\|P\varPsi_1(-1)\|^2}{\mu_1(-1)} = \fraction{\|P^2\varPsi_0(0)\|^2}{\|P\varPsi_0(0)\|^2} = \fraction{1}{2},\quad
    q_1(-1) =
    \fraction{\|Q\varPsi_1(-1)\|^2}{\mu_1(-1)} = \fraction{\|QP\varPsi_0(0)\|^2}{\|P\varPsi_0(0)\|^2} = \fraction{1}{2},\\
    &p_1(1) = \fraction{\|P\varPsi_1(1)\|^2}{\mu_1(1)} = \fraction{\|PQ\varPsi_0(0)\|^2}{\|Q\varPsi_0(0)\|^2} = \fraction{1}{2},\quad
    q_1(1) =
    \fraction{\|Q\varPsi_1(1)\|^2}{\mu_1(1)} = \fraction{\|Q^2\varPsi_0(0)\|^2}{\|Q\varPsi_0(0)\|^2} = \fraction{1}{2}.
\end{align}
Therefore, the existence probabilities at time $n=2$ are obtained as follows:
\begin{align}
    \nu_2(-2) = p_1(-1)\nu_1(-1) = \fraction{1}{4},\quad
    \nu_2(0) = p_1(1)\nu_1(1) + q_1(-1)\nu_1(-1)=\fraction{1}{2},\quad
    \nu_2(2) = q_1(1)\nu_1(1) = \fraction{1}{4}.
\end{align}
The probabilities $p_2(0)$, $q_2(0)$, $p_2(\pm 2)$, $q_1(\pm 2)$ are calculated as follows:
\begin{align}
p_2(-2) = q_2(-2) = p_2(0) = q_2(0) = p_2(2) = q_2(2) = \fraction{1}{2}.
\end{align}
Therefore, the existence probabilities at time $n=3$ are obtained as follows:
\begin{align}
    \nu_3(-3) = \nu_3(3) = \fraction{1}{8},\quad \nu_3(-1) = \nu_3(1) = \fraction{3}{8}.
\end{align}
Until then, the probabilities are the same as the simple random walk, but this shall not apply to further behavior. The transition probabilities at $n=3$ are calculated as follows:
\begin{align}
&p_3(-3) = q_3(-3) = p_3(3) = q_3(3) = \fraction{1}{2},\\
&p_3(-1) = q_3(1) = \fraction{5}{6},\quad q_3(-1) = p_3(1)  = \fraction{1}{6}.
\end{align}
Here we note that $\varPsi_3(-3) = P^3\varPsi_0(0)$, $\varPsi_3(-1) = (P^2Q + PQP +QP^2)\varPsi_0(0)$, $\varPsi_3(1) = (PQ^2 + QPQ + Q^2P)\varPsi_0(0)$, and $\varPsi_3(3) = Q^3\varPsi_0(0)$. Therefore, the existence probabilities at time $n=4$ are obtained as follows:
\begin{align}
	\nu_4(-4) = \nu_4(4) = \fraction{1}{16},\quad 
	\nu_4(-2) = \nu_4(2) = \fraction{3}{8},\quad 
	\nu_4(0) = \fraction{1}{8}.
\end{align}
The probabilities at time $n=5$ is calculated similarly.}\revise{\revCalA}

\revise{\subsection{Ex. 2 [One-defect model]}}\noindent
\def\revCalB{The coin matrix $C_x$ is decomposed as follows:
\begin{align}
	P_x = \left\{\begin{array}{ll}
\twobytwo{1}{0}{0}{0} & (x=0) \vrule width 0pt height 9pt depth 13pt\\
\fraction{1}{\sqrt{2}}\twobytwo{1}{1}{0}{0} =: P& (x\not =0) 
\end{array}\right.,\quad
	Q_x = \left\{\begin{array}{ll}
\twobytwo{0}{0}{0}{1} & (x=0) \vrule width 0pt height 9pt depth 13pt\\
\fraction{1}{\sqrt{2}}\twobytwo{0}{0}{1}{-1} =: Q& (x\not =0) 
\end{array}\right..
\end{align}
Equally to Ex. 1, we define the initial probability distribution as
\begin{align}
	\nu_0(0) = \mu_0(0) = 1.
\end{align}
\begin{align}
    p_0(0) = \fraction{\|P_0\varPsi_0(0)\|^2}{\mu_0(0)}  = \fraction{1}{2},\,\,\,\,
    q_0(0) = \fraction{\|Q_0\varPsi_0(0)\|^2}{\mu_0(0)}  = \fraction{1}{2}.
\end{align}
Therefore, the existence probabilities at time $n=1$ are obtained as follows:
\begin{align}
    \nu_1(-1) = p_0(0)\nu_0(0) = \fraction{1}{2},\,\,\,\,
    \nu_1(1) = q_0(0)\nu_0(0) = \fraction{1}{2}
\end{align}
The probabilities $p_1(\pm 1)$, $q_1(\pm 1)$ are calculated as follows:
\begin{align}
    &p_1(-1) = \fraction{\|P\varPsi_1(-1)\|^2}{\mu_1(-1)} = \fraction{\|PP_0\varPsi_0(0)\|^2}{\|P_0\varPsi_0(0)\|^2} = \fraction{1}{2},\quad
    q_1(-1) =
    \fraction{\|Q\varPsi_1(-1)\|^2}{\mu_1(-1)} = \fraction{\|QP_0\varPsi_0(0)\|^2}{\|P_0\varPsi_0(0)\|^2} = \fraction{1}{2},\\
    &p_1(1) = \fraction{\|P\varPsi_1(1)\|^2}{\mu_1(1)} = \fraction{\|PQ_0\varPsi_0(0)\|^2}{\|Q_0\varPsi_0(0)\|^2} = \fraction{1}{2},\quad
    q_1(1) =
    \fraction{\|Q\varPsi_1(1)\|^2}{\mu_1(1)} = \fraction{\|QQ_0\varPsi_0(0)\|^2}{\|Q_0\varPsi_0(0)\|^2} = \fraction{1}{2}.
\end{align}
The result is the same as Ex. 1, but we should note that the applied matrices are different from Ex. 1 because of the definition of the coin. The existence probabilities at time $n=2$ are obtained as follows:
\begin{align}
    \nu_2(-2) = p_1(-1)\nu_1(-1) = \fraction{1}{4},\quad
    \nu_2(0) = p_1(1)\nu_1(1) + q_1(-1)\nu_1(-1)=\fraction{1}{2},\quad
    \nu_2(2) = q_1(1)\nu_1(1) = \fraction{1}{4}.
\end{align}
Equally, by using the decomposition matrices properly, we obtain the probabilities after then.
}\revise{\revCalB}\vspace{2\baselineskip}

\section{The 2-term moving variance of $\bfit{r(n)}$}
\def\revSec{We define 2-term moving variance $v(n)$ of $r(n)$, which we introduced as Eq. (\ref{eq:mv}) in Sec.~3.2, along 2-term moving average $m(n)$ of $r(n)$ defined as follow.}\revise{\revSec} For $n\in\mathbb{N}$,
\begin{align}\label{mn}
m(n) = \fraction{1}{2}(r(n)-r(n-1)).
\end{align}
Here we consider $v(n)$ as the analog of the ordinary variance: the mean of the difference between the random variable and its mean. Concretely, we define
\begin{align}
v(n) = \fraction{1}{2}\{(r(n) - m(n))^2 + (r(n-1)-m(n))^2\}.
\end{align}
By substituting (\ref{mn}) to it, we obtain
\begin{align}
v(n) = \fraction{1}{4}(r(n)^2-2r(n)r(n-1)+5r(n-1)^2).
\end{align}

\section*{Acknowledgment}
This work was supported by SPRING program (JPMJSP2108) and CREST project and in part by the CREST project (JPMJCR17N2) funded by the Japan Science and Technology Agency and Grants-in-Aid for Scientific Research (JP20H00233) funded by the Japan Society for the Promotion of Science.

\section*{Data availability statement}
The datasets generated during and/or analyzed during the current study are available from the corresponding author on reasonable request.

\bibliographystyle{ieeetr}
\bibliography{Bib_qr}
\end{document}